\documentclass{article}
\usepackage{spconf,amsmath,graphicx, amssymb, hyperref}
\usepackage{multirow, xcolor, soul, enumitem}
\usepackage[linesnumbered,lined,boxed,commentsnumbered]{algorithm2e}

\RestyleAlgo{ruled}

\newcommand{\vZ}{\mathbf{Z}}
\newcommand{\vX}{\mathbf{X}}
\newcommand{\vY}{\mathbf{Y}}

\def\x{{\mathbf x}}
\def\y{{\mathbf y}}
\def\z{{\mathbf z}}
\def\eps{{\boldsymbol \epsilon}}

\title{CONDITIONING AND SAMPLING IN VARIATIONAL DIFFUSION MODELS FOR SPEECH SUPER-RESOLUTION}

\name{Chin-Yun Yu$^{1,}$\sthanks{The majority of this work was conducted independently before the author enrolled in Queen Mary University of London.}, Sung-Lin Yeh$^2$, Gy{\"o}rgy Fazekas$^1$, Hao Tang$^2$}
\address{$^1$Centre for Digital Music, Queen Mary University of London, UK\\
$^2$Institute for Language, Cognition and Computation, University of Edinburgh, UK}

\begin{document}
%
\maketitle
\begin{abstract}
Recently, diffusion models (DMs) have been increasingly used in audio processing tasks, including speech super-resolution (SR), which aims to restore high-frequency content given low-resolution speech utterances.
This is commonly achieved by conditioning the network of noise predictor with low-resolution audio.
In this paper, we propose a novel sampling algorithm that communicates the information of the low-resolution audio via the reverse sampling process of DMs.
The proposed method can be a drop-in replacement for the vanilla sampling process and can significantly improve the performance of the existing works.
Moreover, by coupling the proposed sampling method with an unconditional DM, i.e., a DM with no auxiliary inputs to its noise predictor, we can generalize it to a wide range of SR setups.
We also attain state-of-the-art results on the VCTK Multi-Speaker benchmark with this novel formulation.

\end{abstract}
\section{Introduction}
\label{sec:intro}

Speech super-resolution (SR), also known as bandwidth extension, is a speech generation task that reconstructs high-resolution speech utterances from low-resolution ones.
NU-Wave~\cite{leeNUWaveDiffusionProbabilistic2021} and NU-Wave 2~\cite{hanNUWaveGeneralNeural2022} are based on diffusion models (DMs)~\cite{sohl2015deep, songSCOREBASEDGENERATIVEMODELING2021, hoDenoisingDiffusionProbabilistic2020, chenWaveGradEstimatingGradients2020, kongDiffWaveVersatileDiffusion2021,  kingmaVariationalDiffusionModels2022} and are able to reconstruct speech up to 48 kHz. 
However, due to the stochastic nature of the diffusion sampling, and because low-resolution audio is only fed into the noise predictor network, the low-frequency spectra between the reconstructed output and the input can deviate~\cite{hanNUWaveGeneralNeural2022}, where the difference should ideally be zero.
We aim to explore the possibility to enhance existing diffusion SR models.

DMs are not only impressive generative models, but also powerful unsupervised problem solvers, by incorporating the condition into the reverse sampling process with unconditional DMs~\cite{songSCOREBASEDGENERATIVEMODELING2021, kawarDenoisingDiffusionRestoration2022, lugmayrRePaintInpaintingUsing2022, chungImprovingDiffusionModels2022, songSOLVINGINVERSEPROBLEMS2022}.
This framework has shown impressive results in various inverse problems for image processing, such as image inpainting~\cite{kawarDenoisingDiffusionRestoration2022, lugmayrRePaintInpaintingUsing2022, chungImprovingDiffusionModels2022}, image super-resolution~\cite{kawarDenoisingDiffusionRestoration2022, chungImprovingDiffusionModels2022}, and medical image reconstruction~\cite{songSOLVINGINVERSEPROBLEMS2022}. 
For speech processing, although DMs have been used as vocoders~\cite{chenWaveGradEstimatingGradients2020, kongDiffWaveVersatileDiffusion2021}, speech enhancement~\cite{luStudySpeechEnhancement2021, welkerSpeechEnhancementScoreBased2022}, and speech SR~\cite{leeNUWaveDiffusionProbabilistic2021, hanNUWaveGeneralNeural2022}, there has been no similar attempt to design a sampling method for recovering audio signals beyond additive noise~\cite{luStudySpeechEnhancement2021, welkerSpeechEnhancementScoreBased2022}.

In this work, we cast the task of speech SR as an inpainting problem in the frequency-domain, and propose a diffusion sampling algorithm.
The proposed method can be a drop-in replacement for the reverse sampling process in other diffusion SR models for quality improvements.
Moreover, by combining the method with a UDM, our approach can generalize to various SR conditions, such as varying upscaling ratios and types of downsampling filters.
We also observed that training audio DMs on the variational lower bound is beneficial for our method.
Finally, we outperform the state-of-the-art in terms of \emph{log-spectral-distance} (LSD) on the 48kHz VCTK Multi-Speaker benchmark, demonstrating robustness against different downsampling setups.\footnote{The source code and audio samples are available from \url{https://iamycy.github.io/diffwave-sr}.}

\section{Background}
\label{sec:back}

\subsection{Audio resampling}
\label{ssec:ar}
Audio resampling is a function that maps the discrete audio signal $\x = [x_1, x_2,...,x_{Tl}]$ sampled at frequency $l$ to $\y = [y_1, y_2,...,y_{Th}]$ with a new sampling frequency $h$, where $T$ is the length in seconds. It can further be categorized as downsampling or upsampling depending on whether $l$ is greater or smaller than $h$. According to the Nyquist theorem, the highest frequency a signal can contain is half of its sampling frequency. Thus, a lowpass filter with a cutoff frequency $\min(l, h)/2$ is applied to avoid aliasing when resampling $\x$. 
Notice that downsampling eliminates all frequency information between $l/2$ to $h/2$. 
Reversing this process is the essence of SR.
To distinguish it from regular upsampling, we will refer to this operation as upscaling and use upscaling ratio for $l/h$ in later paragraphs.

\subsection{Variational diffusion models}
\label{ssec:vdm}

Variational DMs (VDMs)~\cite{sohl2015deep, hoDenoisingDiffusionProbabilistic2020, kingmaVariationalDiffusionModels2022} can generate complex samples starting from a simple normal distribution with a Markov chain structure. In particular, VDM assumes an observed sample $\x$ is generated by a chain of $T$ latent variables $\z_t$. Sampling $\z_t$ from $\x$ is trivial because its distribution is parameterized as
\begin{equation}
\label{eq:latent}
q(\z_t|\x) = \mathcal{N}(\alpha_t \x, \sigma_t^2 \mathbf{I}),
\end{equation}
where $\alpha_t \in (0, 1)$, $\sigma_t^2 = 1 - \alpha_t^2$, $1 \leq t \leq T$. $\alpha_t$ is monotonically decreasing in $t$, making $\z_t$ noisier in $t$ and eventually $q(\z_T|\x) \approx \mathcal{N}(0, \mathbf{I})$. Sampling $\x$, however, requires a chain of actions. Starting from $\z_T$ sampled from $p(\z_T) = \mathcal{N}(0, \mathbf{I})$, the \emph{reverse diffusion} step $p(\z_{t-1}|\z_t)$ is applied repeatedly until $\z_1 \approx \x$. We do not know the distribution of $p(\z_{t-1}|\z_t)$, but we can approximate it by maximizing the variational lower bound (VLB)
\begin{equation}
\label{eq:vlb}
\begin{split}
\log p(\x) \geq VLB(\x) &= \mathbb{E}_{q(\z_1|\x)}[\log p(\x|\z_1)] \\
& - D_{KL}(q(\z_T|\x) || p(\z_T)) \\
& - \mathcal{L}_T(\x),
\end{split}
\end{equation}
with diffusion loss
\begin{equation}
\label{eq:diff_loss}
\mathcal{L}_T(\x) = \sum_{t=2}^T \mathbb{E}_q \left[ D_{KL} \left (q(\z_{t-1}|\z_{t}, \x) || p(\z_{t-1}|\z_{t}) \right) \right].
\end{equation}
The term $q(\z_{t-1}|\z_t, \x)$ can be derived from Eq.~\eqref{eq:latent} using Bayes rule and is given by
\begin{equation}
\label{eq:rev_x}
\begin{split}
& q(\z_{t-1}|\z_t, \x) = \\
\mathcal{N} &
\left(
\frac{\alpha_{\bar{t}}\sigma_{t-1}^2}{\sigma_t^2}\z_t
+ \frac{\alpha_{t-1}\sigma_{\bar{t}}^2}{\sigma_t^2}\x
, \frac{\sigma_{\bar{t}}^2\sigma_{t-1}^2}{\sigma_t^2}\mathbf{I}
\right),
\end{split}
\end{equation}
where $\alpha_{\bar{t}} = \alpha_t/\alpha_{t-1}$ and $\sigma_{\bar{t}}^2 = \sigma_t^2 - \alpha_{\bar{t}}^2\sigma_{t-1}^2 $. To train the model, a common practice is to parameterize $p(\z_{t-1}|\z_t)$ as $q(\z_{t-1}|\z_t, \x=\hat{\x}_\theta(\z_t;t))$ where $\hat{\x}_\theta(\z_t;t) = (\z_t - \sigma_t \hat{\eps}_\theta(\z_t;t))/\alpha_t$ and $\hat{\eps}_\theta(\z_t;t)$ is a neural network that predicts the added noise from $\z_t$~\cite{hoDenoisingDiffusionProbabilistic2020, kingmaVariationalDiffusionModels2022}. 

For training efficiency, we can use the following equation to replace Eq.~\eqref{eq:diff_loss}:
\begin{equation}
\label{eq:est_diff_loss}
\mathcal{L}_\infty(\x) = 
\frac{\delta_{max} - \delta_{min}}{2}\mathbb{E}_{\eps, \upsilon} \left[ \|\eps - \Tilde{\eps}_\theta(\z_\upsilon; \upsilon) \|_2^2 \right],
\end{equation}
where $\delta_{max} = \log (\alpha_1^2/\sigma_1^2), \delta_{min} = \log (\alpha_\infty^2/\sigma_\infty^2), \eps \sim \mathcal{N}(0, \mathbf{I}), \upsilon \sim \mathcal{U}(\delta_{min}, \delta_{max})$, and $\z_\upsilon = \sqrt{\phi(\upsilon)}\x + \sqrt{\phi(-\upsilon)}\eps$, $\phi$ is the sigmoid function. 
Eq.~\eqref{eq:est_diff_loss} is a stochastic estimate of diffusion loss when $T \rightarrow \infty$.
The related proof can be found in~\cite{kingmaVariationalDiffusionModels2022}.
Prior works of audio DMs~\cite{leeNUWaveDiffusionProbabilistic2021, hanNUWaveGeneralNeural2022, chenWaveGradEstimatingGradients2020, kongDiffWaveVersatileDiffusion2021, luStudySpeechEnhancement2021, welkerSpeechEnhancementScoreBased2022} were trained with a biased diffusion loss~\eqref{eq:diff_loss} and fixed noise schedule endpoints ($\alpha_1$ and $\alpha_\infty)$.
However, Kingma et al.~\cite{kingmaVariationalDiffusionModels2022} show that training on the VLB with a learnable $\alpha_t$ can improve the likelihood on unconditional generation.

\section{Approach}
\label{sec:approach}
Assuming we have $\alpha_t$ and $\hat{\eps}_\theta(\z_t;t)$ that accurately estimate the high-resolution audio distribution $p(\x)$, we can then leverage it to perform audio super-resolution by injecting low-resolution audio $\y$ into the sampling process if the downsampling scheme is known. 
In other words, we can transform the prior distribution $p(\x)$ into a conditional distribution $p(\x|\y)$ by replacing $p(\z_{t-1}|\z_t)$ with the \emph{conditional reverse diffusion} step $p(\z_{t-1}|\z_t, \y)$.
We introduce $\hat{\y}$ by upsampling $\y$ to make sure all the signals have the same dimension. 
The process of downsampling followed by upsampling can be viewed as a low pass filter $\mathcal{F}_{h/2}$ with cut off frequency $h/2$ so $\hat{\y} = \mathcal{F}_{h/2}(\x)$.
Let $\vX^\omega, \hat{\vY}^\omega, \vZ_t^\omega, \hat{\vX}_t^\omega \in \mathbb{C}$ be the response of $\x, \hat{\y}, \z_t, \hat{\x}_\theta(\z_t; t)$ at frequency $\omega$ after a normalized Fourier transform. 
We can rewrite Eq.~\eqref{eq:rev_x} in the frequency-domain as
\begin{equation}
\begin{split}
& q(\vZ_{t-1}^\omega|\vZ_t^\omega, \vX^\omega) = \\
\mathcal{CN} &
\left(
\frac{\alpha_{\bar{t}}\sigma_{t-1}^2}{\sigma_t^2}\vZ_t^\omega
+ \frac{\alpha_{t-1}\sigma_{\bar{t}}^2}{\sigma_t^2}\vX^\omega
, \frac{\sigma_{\bar{t}}^2\sigma_{t-1}^2}{\sigma_t^2}
\right).
\end{split}
\end{equation}
$\hat{\vY}^\omega$ equals $c_\omega\vX^\omega$ where $c_\omega$ is the coefficient of $\mathcal{F}_{h/2}$ at frequency $\omega$. 
If $\mathcal{F}_{h/2}$ is ideal, $c_\omega$ equals one when $\omega < h/2$ and zero otherwise. 
In order to utilize the available $\vX^\omega$ from $\hat{\vY}^\omega$, we propose to parameterize the reverse step as
\begin{equation}
\label{eq:rev_cond}
\begin{split}
&p(\vZ_{t-1}^\omega|\vZ_t^\omega, \hat{\vY}^\omega) = \\
&q(\vZ_{t-1}^\omega|\vZ_t^\omega, \vX^\omega=\hat{\vY}^\omega + (1 - c_\omega)\hat{\vX}_t^\omega).
\end{split}
\end{equation}
What Eq. \eqref{eq:rev_cond} does is it replaces $\hat{\vX}_t^\omega$ in the low-frequency region with ground truth $\vX^\omega$.
In this way, we can sample low frequencies from the ground truth information, and the deviation of low frequencies in the final output $\hat{\x}_\theta(\z_1; 1)$ is then proportional to the variance $\sigma_1^2/\alpha_1^2$ according to Eq. \eqref{eq:latent}. 
Although we still use the estimation $\hat{\vX}_t^\omega$ to sample high frequencies, since $\hat{\eps}_\theta(\z_t; t)$ utilize all the frequency in $\z_t$ for noise prediction, the low-frequency information of $\x$ in $\z_t$ can implicitly guide the whole process to gradually \emph{inpaint} the missing high frequencies that are coherent to the provided low frequencies.
In addition, $\hat{\eps}_\theta(\z_t; t)$ can come from any DM which is either unconditional or only conditioned on $\y$. 

Most of the existing audio DMs operate in the time-domain~\cite{leeNUWaveDiffusionProbabilistic2021, hanNUWaveGeneralNeural2022, chenWaveGradEstimatingGradients2020, kongDiffWaveVersatileDiffusion2021, luStudySpeechEnhancement2021}. 
Fortunately, the time-domain version of Eq.~\eqref{eq:rev_cond}, $p(\z_{t-1}|\z_t, \hat{\y})$, is equal to $q(\z_{t-1}|\z_t, \x=\hat{\y} + \hat{\x}_\theta(\z_t;t) - \mathcal{F}_{h/2}(\hat{\x}_\theta(\z_t;t)))$. 
If we use time-domain filters for $\mathcal{F}_{h/2}$ then the whole reverse process can be done in the time-domain.
On the one hand, this process is similar to RePaint~\cite{lugmayrRePaintInpaintingUsing2022} but the masking is performed in the frequency-domain based on Eq. \eqref{eq:rev_cond}.
On the other hand, the use of filters helps to perform diffusion without projecting $\z_t$ into spectral space in every iteration, either using transformation matrices calculated from SVD~\cite{kawarDenoisingDiffusionRestoration2022} or Fourier transform~\cite{songSOLVINGINVERSEPROBLEMS2022}, thus it is computationally more efficient. 
In practice, we set $p(\x|\z_1, \y)$ the same as $p(\x|\z_1)$ and assume $p(\z_T|\hat{\y}) = p(\z_T)$.

To further increase performance, we also incorporate the Manifold Constrained Gradient (MCG)~\cite{chungImprovingDiffusionModels2022} method into our algorithm. Specifically, we subtract a gradient term $\partial \|\hat{\y} - \mathcal{F}_{h/2}(\hat{\x}_\theta(\z_t;t)))\|_2^2 / \partial \z_t$ in each reverse step. 
We control this gradient term with a step size $\eta$.
The full inference procedure is summarized in Algorithm \ref{alg:sampling}.

\begin{algorithm}
\DontPrintSemicolon 
\caption{Conditional sampling}\label{alg:sampling}
\SetAlgoNlRelativeSize{0}
\SetNlSty{}{}{:}
\SetKwData{Left}{left}\SetKwData{This}{this}\SetKwData{Up}{up}
\SetKwFunction{Upsample}{Upsample}\SetKwFunction{Downsample}{Downsample}
\SetKwComment{Comment}{\% }{}
\SetKwProg{Def}{def}{:}{}
\SetKwInOut{Require}{Require}\SetKwInOut{Output}{output}

\Require{$\y, T, \eta$}

\Def{$\mathcal{F}_{h/2}(\x)$}{\Return \Upsample{\Downsample{$\x$}}}
$\hat{\y} \leftarrow$ \Upsample{$\y$}\;
$\z_T \sim \mathcal{N}(0, \mathbf{I})$\;
\For {$t = T, T-1,...,2$}{
    $\hat{\x} \leftarrow (\z_t - \sigma_t \tilde{\eps}_\theta(\z_t; \delta_t))/\alpha_t$\;
    \textcolor{orange}{$\mathbf{g} \leftarrow \frac{\partial}{\partial \z_t}  \|\hat{\y} - \mathcal{F}_{h/2}(\hat{\x}) \|_2^2$  \Comment*[r]{MCG}}
    \textcolor{blue}{$\hat{\x} \leftarrow \hat{\y} + \hat{\x} - \mathcal{F}_{h/2}(\hat{\x})$  \Comment*[r]{Inpainting}}
    $\boldsymbol{\mu} \leftarrow \frac{\alpha_{\bar{t}}\sigma_{t-1}^2}{\sigma_t^2}\z_t
                      + \frac{\alpha_{t-1}\sigma_{\bar{t}}^2}{\sigma_t^2}\hat{\x}$\;
    \textcolor{orange}{$\boldsymbol{\mu} \leftarrow \boldsymbol{\mu} - \eta (\mathbf{g} - \mathcal{F}_{h/2}(\mathbf{g}))$  \Comment*[r]{MCG update}}
    $\eps \sim \mathcal{N}(0, \mathbf{I})$\;
    $\z_{t-1} \leftarrow \boldsymbol{\mu}+\frac{\sigma_{\bar{t}}^2\sigma_{t-1}^2}{\sigma_t^2}\eps$\;
}
$\hat{\x} \leftarrow (\z_1 - \sigma_1 \tilde{\eps}_\theta(\z_1; \delta_{max}))/\alpha_1$ \;
\Return $\hat{\x}$
\end{algorithm}

\begin{table*}[t]
\centering
\caption{A list of models we used in the paper and their characteristics. 
By unconditional we mean that the sampling algorithm does not involve condition signals. 
The underlined variables are only used at the inference stage.}
\label{table:models}
\begin{tabular}{cccccc}
\hline
\textbf{Model}    & \textbf{Modeling}  & \textbf{Architecture}   & \textbf{Inputs}   & \textbf{Loss Function}           & \textbf{Sampling} \\ \hline
NU-Wave   & Diffusion & DiffWave                     & $\z_t, \y$                                                   & Biased $\mathcal{L}_T(\x)$   & Uncond.    \\
NU-Wave 2 & Diffusion & Modified DiffWave            & $\z_t, \y, h/2$                                              & Biased $\mathcal{L}_T(\x)$     & Uncond.    \\
WSRGlow   & Flow      & WaveGlow                     & $\y, \underline{\eps} \sim \mathcal{N}(0, \mathbf{I})$       & $-\log p(\x)$           & Uncond.    \\
NVSR      & Various         & UNet based                   & $\y$                                                         & L1 Loss                   & Deterministic                \\
UDM+      & Diffusion & DiffWave w/o Cond. Layers &  $\z_t, \underline{\y}$                                        & $-VLB(\x)$                    & Cond.     \\ \hline
\end{tabular}
\end{table*}

\section{Experiments}
\label{sec:exp}

\subsection{Dataset}
\label{ssec:dataset}
We run all the experiments on the VCTK-0.92 dataset~\cite{vctk092} which contains 44 hours of speech recordings spoken by 108 English speakers with different accents. 
We follow the same  configurations proposed in~\cite{leeNUWaveDiffusionProbabilistic2021} to split training and test data. 
We did the evaluations at the 48 kHz and 16 kHz sampling rate. 
To generate low-resolution audio, we consider two configurations of low pass filter for downsampling: 1) the spectrogram-based method following the settings proposed in~\cite{leeNUWaveDiffusionProbabilistic2021} and 2) sinc filter method from~\cite{RESAMPLE}.
These filters all have narrow passband and zero phase response~\footnote{The settings for the sinc filter are 128 zero-crossings and 0.962 roll-off, and we windowed it using a Kaiser window with $\beta \approx 14.77$.}, serving as good approximations to the ideal filter.

\subsection{Baselines}
\label{ssec:baselines}
For comparison, we choose cubic spline interpolation, NU-Wave~\cite{leeNUWaveDiffusionProbabilistic2021}, NU-Wave 2~\cite{hanNUWaveGeneralNeural2022}, WSRGlow~\cite{zhangWSRGlowGlowbasedWaveform2021}, and NVSR~\cite{liuNeuralVocoderAll2022} as our baselines. We used community-contributed checkpoints\footnote{\url{https://github.com/mindslab-ai/nuwave/issues/18}} of NU-Wave from their official repository.
We trained NU-Wave 2 using the official implementation. 
We reproduced WSRGlow using an open-source implementation\footnote{\url{https://doi.org/10.5281/zenodo.4353123}} of WaveGlow~\cite{prenger2019waveglow} and trained it with original configurations and the spectrogram-based downsampling filter. 
For NVSR we use their official checkpoint.
All baselines were trained on the same training set in Section \ref{ssec:dataset}.
Table \ref{table:models} gives a comparison of the baselines.

\subsection{Model and training configurations}
For the choice of UDM, we adopt DiffWave~\cite{kongDiffWaveVersatileDiffusion2021} and remove all the condition layers. 
We set the channel size to 64 for the 48 kHz model and 128 for the 16 kHz model.
Both UDMs were trained on the negative VLB~\eqref{eq:vlb} for 500k steps using Adam~\cite{kingmaAdamMethodStochastic2017} with a learning rate of 0.0002. 
We parameterize $p(\x|\z_1)$ as $\mathcal{N}(\z_1/\alpha_1, \exp(-\delta_{max}) \mathbf{I})$. 
We initialized $\delta_{max}$ and $\delta_{min}$ to 10 and 0, respectively. 
The model weights are the exponential moving averages of parameters with a 0.9999 momentum. 
We set $\hat{\eps}_\theta(\z_t;t) = \Tilde{\eps}_\theta(\z_{\delta_t}; \delta_t)$ where $\delta_t$ is the \emph{log-signal-to-noise-ratio} scheduler. 
We found that simple linear schedule $\delta_t = \frac{t-1}{T-1} \delta_{min} + \frac{T-t}{T-1}\delta_{max}$ can generate much more consistent samples than other schedulers so we used it in all our experiments. 
We evaluate three variations of our algorithm: 
1) Algorithm \ref{alg:sampling} with our trained UDM as noise predicotr (UDM+), 
2) Algorithm \ref{alg:sampling} with the same noise predictor from the NU-Wave baseline (NU-Wave+), and
3) Algorithm \ref{alg:sampling} with the same noise predictor from the NU-Wave 2 baseline (NU-Wave 2+).
We upsample the input audio to the target resolution using the same sinc filter from Section \ref{ssec:dataset} for all the variations.

\subsection{Evaluation}
We choose the commonly used LSD~\cite{leeNUWaveDiffusionProbabilistic2021, liuNeuralVocoderAll2022, zhangWSRGlowGlowbasedWaveform2021} for objective evaluation and a perceptual metric PESQ~\cite{rix2001perceptual, wang2021towards} for subjective evaluation. 
We evaluate the metrics on each utterance and report the average on the test set. 
For all the diffusion based models, we use the scheduler trained at the same operating frequency for fair comparisons.
We tuned $\eta$ for each evaluation setting empirically using a few utterances from the training set. 
We report all the scores with $T=50$ unless we mention specifically.
We apply downsampling on the model outputs if the model's operating rate is higher than that of the experiment. 
We also compute the LSD below $h/2$ Hz (LSD-LF) in the 48k Hz experiment using simulated input, which is generated by adding random noise with a variance of $\sigma_1^2/\alpha_1^2$ to the test samples.

\section{Results and analysis}
Table \ref{table:vctk48-comparison} shows that our method can enhance existing diffusion baselines, especially for NU-Wave+, which shows a 0.11 to 0.13 reduction in LSD compared to NU-Wave. 
UDM+ outperforms the baselines in all the settings except for ratio of 3 with STFT filter. 
We think the extra gains come from the training objective because this is the only differences between our model and NU-Wave besides the absence of condition layers.
For the 16 kHz experiment, Table \ref{table:vctk16-comparison} shows UDM+ achieves comparable performance to NVSR in LSD.
The inferior result in PESQ can be improved if we remove MCG.
Nevertheless, removing MCG can results in great deterioration in LSD, which can be seen in Table \ref{table:vctk16-comparison} and Table \ref{table:4x-comparison}, especially when using fewer diffusion steps.

The robustness to different filters can also be inspected from the LSD scores in Table \ref{table:vctk48-comparison} and Table \ref{table:vctk16-comparison}. 
UDM+ and NVSR have the smallest differences across filters among all the models. 
Good generalizability to different filters can be achieved through data augmentation~\cite{hanNUWaveGeneralNeural2022, wang2021towards} but our method overcomes this by disentangling the filter conditions from training.
NVSR uses reconstructed mel-spectrograms to generate high-resolution audio, and this low-dimensional representation can blur out the characteristics of the low pass filter. 
The same robustness cannot be achieved if the noise predictor was trained with condition as auxiliary input, as shown in the results from NU-Wave+/NU-Wave 2+. 

Finally, the LSD-LF we obtain for ratio of 2 and 3 are 0.056 and 0.052, respectively, which are roughly six times smaller than the numbers reported from NU-Wave 2~\cite{hanNUWaveGeneralNeural2022}, demonstrating its ability to better retain low frequencies.
In addition, we found our UDMs performed poorly on unconditional generation.
Because our method depends on the quality of prior models, there is still room for improvement.

\begin{table}[t]
\centering
\caption{Evaluation results of LSD($\downarrow$) at 48 kHz target frequency with different upscaling ratios and filter settings.}
\label{table:vctk48-comparison}
\begin{tabular}{ccccc}
\hline
\multirow{2}{*}{\textbf{Model}} & \multicolumn{2}{c}{2x} & \multicolumn{2}{c}{3x} \\ \cline{2-5} 
                                & Sinc      & STFT       & Sinc      & STFT       \\ \hline
Unprocessed                     & 2.81      & 2.80       & 3.18      & 3.18       \\
Spline                          & 2.32      & 2.24       & 2.78      & 2.73       \\
NU-Wave                         & 0.87      & 0.85       & 1.00      & 0.99       \\
NU-Wave 2                       & 0.75      & 0.71       & 0.89      & 0.86         \\
WSRGlow                         & 0.72      & 0.71       & 0.80      & \textbf{0.79}\\
\hline
NU-Wave+                        & 0.74      & 0.72       & 0.89      & 0.88        \\
NU-Wave 2+                      & 0.74      & 0.71       & 0.89      & 0.86         \\
UDM+                            & \textbf{0.64}  & \textbf{0.64} & \textbf{0.79} & \textbf{0.79}    \\ \hline

\end{tabular}
\end{table}

\begin{table}[t]
\centering
\caption{8 kHz to 16 kHz evaluation results with different filter settings.}
\label{table:vctk16-comparison}
\begin{tabular}{ccccc}
\hline
\multirow{2}{*}{\textbf{Model}} & \multicolumn{2}{c}{LSD($\downarrow$)}                & \multicolumn{2}{c}{PESQ($\uparrow$)}   \\ \cline{2-5} 
                                & Sinc                   & STFT                        & Sinc                   & STFT          \\ \hline
Unprocessed                     & 2.56                   & 2.54                        & 1.17                   & 1.17          \\
Spline                          & 1.83                   & 1.72                        & 3.43                   & 3.44          \\
NU-Wave 2                       & 1.06                   & 0.94                        & 3.38                   & 3.33          \\
NVSR                            & 0.81                   & 0.80                        & \textbf{3.47}          & \textbf{3.50} \\ \hline
NU-Wave 2+                      & 0.99                   & 0.90                        & 3.27                   & 3.24          \\
UDM+                            & \textbf{0.79}          & \textbf{0.78}               & 2.93                   & 2.96          \\ 
NU-Wave 2+ w/o MCG              & 1.04                   & 0.94                        & 3.46                   & 3.39          \\
UDM+ w/o MCG                    & 0.90                   & 0.90                        & 3.43                   & 3.44          \\ \hline
\end{tabular}
\end{table}

\begin{table}[]
\centering
\caption{Evaluation results of LSD($\downarrow$) at 48 kHz target frequency with a upscaling ratio of 4 and different number of diffusion steps.}
\label{table:4x-comparison}
\begin{tabular}{ccccc}
\hline
\textbf{Model}        & T = 25 & T = 50 & T = 100 & T = 200 \\ \hline
UDM+ w/o MCG          & 1.11   & 0.96   & 0.88    &  0.86        \\
UDM+                  & 0.92   & 0.90   & 0.86    &  \textbf{0.84}   \\ \hline
WSRGlow      &  \multicolumn{4}{c}{ 0.85}          \\ \hline
\end{tabular}
\end{table}

\section{Conclusion}
\label{sec:conclu}
We presented a conditional sampling algorithm for SR that can be flexibly coupled with existing audio DMs.
Our experiments on the 48 kHz SR benchmark showed that our conditional sampling with MCG improved the performance of previous conditional audio DMs and achieve state-of-the-art when combined with a UDM, which was trained on the unbiased diffusion loss.
We also demonstrated the robustness of our method against different zero-phase downsampling filters.

\section{Acknowledgement}
\label{sec:ack}
We would like to thank Yin-Jyun Luo, Sungkyun Chang, and Christian James Steinmetz for giving feedback when drafting the paper.
The first author is a research student at the UKRI CDT in AI and Music, supported jointly by UK Research and Innovation [grant number EP/S022694/1] and Queen Mary University of London.

\bibliographystyle{IEEEbib}
\bibliography{refs}

\end{document}